\documentstyle{article}

\def\ie{\hbox{\it i.e.}{}}      
\def\eg{\hbox{\it e.g.}{}}      \def\cf{\hbox{\it cf.}{}}

\font\tenbf=cmbx10
\font\tenrm=cmr10
\font\tenit=cmti10
\font\elevenbf=cmbx10 scaled\magstep 1
 1
\font\elevenit=cmti10 scaled\magstep 1

\renewenvironment{thebibliography}[1]
{   \begin{list}{\arabic{enumi}.}
    {\usecounter{enumi} \setlength{\parsep}{0pt}
     \setlength{\itemsep}{2pt} \settowidth{\labelwidth}{#1.}
     \sloppy
    }}{\end{list}}
\begin{document}
\newcommand{\cm}{Commun.\ Math.\ Phys.~}
\newcommand{\pr}{Phys.\ Rev.\ D~}
\newcommand{\pl}{Phys.\ Lett.\ B~}
\newcommand{\np}{Nucl.\ Phys.\ B~}

\begin{flushright}
NUB--3059\\[1mm]
November 1992
\end{flushright}
\vskip 1cm

\begin{center}
{\tenbf LOW-ENERGY EFFECTIVE ACTION OF SUPERSTRING THEORY\\}
\vglue 5pt
\vskip 0.4in
{\tenrm T.R. TAYLOR\\}
\baselineskip=13pt
{\tenit Department of Physics, Northeastern University\\}
\baselineskip=12pt
{\tenit  Boston, MA 02115, U.S.A.\\}
\vglue 0.8cm
{\tenrm ABSTRACT}
\end{center}
\vglue 0.3cm
{\rightskip=3pc
 \leftskip=3pc
 \tenrm\baselineskip=12pt
 \noindent
A fundamental task for the heterotic superstring theory
is the determination of the effective action
describing the physics of massless string excitations
at low energies. This is necessary for the phenomenological
applications of string theory, in particular for the unification
of gauge interactions and for the gaugino condensation mechanism
of supersymmetry breaking. In this talk, I report on the recent
progress in computing the effective supergravity action
from  superstring scattering amplitudes,
at the tree level and beyond.  I discuss
moduli-dependent string loop corrections to gauge and Yukawa couplings.
\vskip 5mm
\begin{center}
\it Talk presented at the 7th Meeting of the American Physical Society\\
Division of Particles and Fields, Fermilab, November 10-14, 1992
\end{center}

\vglue 1cm}
{\elevenbf\noindent 1. Introduction}
\vglue 0.2cm

The basic property of string theory, which makes it so attractive from
the point of view of particle physics, is that the physical couplings
and masses are in principle calculable. They are determined by
the vacuum expectation values (VEVs) of massless scalar fields,
like the dilaton and moduli. Any serious attempt to compute
the low-energy parameters from string theory must address two
basic questions: 1) How do masses and couplings depend on
the VEVs of dilaton, moduli, Higgs scalars {\em etc}.? and
2) What fixes these VEVs?
In the past several years, there has been some slow but steady progress
towards answering the second question; however, the general
perception is that the problem of scalar VEVs still escapes a satisfactory
solution.
Here, I have nothing new to say about this problem.
What is however very surprising, taking into account this stalemate in
superstring phenomenology, is the amount of progress
regarding the first question, of the determination of the
dependence of physical quantities on scalar VEVs, which I will
generically call the problem of moduli-dependence of physical parameters.

A very efficient method of studying the moduli-dependence of low-energy
parameters, \linebreak developed in the last couple of years,
relies on the computation of the effective supergravity action describing the
physics of massless string excitations. The moduli dependence
of the effective action can be
determined by evaluating the appropriate superstring scattering
amplitudes.$^{1,2,3,4}$
In this process, supergravitational interactions are
determined directly from superstring theory.
The moduli-dependent loop corrections obtained in this way give rise to
the so-called threshold corrections to superstring unification parameters,$^5$
and may be phenemenologically relevant in the future, once one understands
the mechanism that fixes the moduli VEVs.
They may also be relevant for the gaugino condensation mechanism
of supersymmetry breaking.
In this talk, I will discuss the structure of low-energy
supergravity that emerges from direct string computations of the tree-level
and one-loop scattering amplitudes.
I will also report the
results of recent computations of threshold corrections
to Yukawa couplings.$^4$

\vglue 0.4cm
{\elevenbf\noindent 2. Superstring Supergravity}
\vglue 0.2cm

The massless spectrum of heterotic superstring theory contains
the supergravity multiplet, gauge multiplets, and a large number of chiral
multiplets. In addition, there is a dilaton which belongs to a
very distinct supersymmetry multiplet, together with the two-index
antisymmetric tensor -- the Kalb-Ramond field. The dilaton VEV plays the role
of
the string loop expansion parameter. Since the Kalb-Ramond field is equivalent
to a pseudoscalar axion, one usually represents the dilaton and its
supersymmetric partners by one chiral multiplet $S$.
\vglue 0.2cm
{\elevenit\noindent 2.1. Tree Level}
\vglue 0.1cm

The most general $N=1$ supergravity action, describing local interactions
involving up to two derivatives, is characterized by three functions
of chiral superfields: the K\"ahler potential $K$ which determines
the kinetic terms, the analytic superpotential $W$ related to the
Yukawa couplings, and the analytic function $f$ associated with the gauge
couplings. At the tree level, the general stucture of the $f$-function
and the K\"ahler potential is common to all compactifications:$^6$
\begin{equation}
f^{(0)}=kS~,~~~~~~~
K^{(0)}=-\ln (S+\,\overline{\! S})+G^{(0)}(Z,\,\overline{\! Z})\, ,
\label{tree}\end{equation}
where $k$ is the level of the Ka\v{c}-Moody algebra
that generates the gauge group, and
$Z$ denote chiral superfields other than $S$. The function
$G^{(0)}(Z,\,\overline{\! Z})$ as well as the superpotential $W(Z)$
depend on the details of compactification and can be determined by using
a number of different methods. The method that I will discuss here is
is based on direct computation of superstring scattering amplitudes.
A typical amplitude that allows determination of the tree-level
K\"ahler potential involves four scalar fields.
Such amplitudes have been explicitly computed at the tree level
for orbifold compactifications.$^1$

\underline{\em Example}.
As an example, consider a class of orbifolds with the gauge group
$E_8\otimes E_6\otimes U(1)^2$, $k=1$,
which contain three untwisted families of 27's,
$A_j$, $j=1,2,3$, in one-to-one correspondence
with the three untwisted moduli $T_j$. In this case,\vglue -3mm
\begin{equation}
W=A_1A_2A_3+\dots~,~~~~~~G^{(0)}=-\sum_{j=1}^3\ln
(T_j+\overline{T}_{\bar{\jmath}}
- A_j\bar{A}_{\bar{\jmath}})+\dots
\label{ex1}\end{equation}
\vglue -0.3cm
{\elevenit\noindent 2.2. One Loop and Beyond}
\vglue 0.1cm

Since the dilaton VEV plays the role of the role of the string loop expansion
parameter, string loop corrections give rise to kinetic terms
that mix the dilaton with the moduli.
This complicates
the analysis of the moduli-dependence of the effective action.
The problem can be avoided by representing the dilaton
and its supersymmetric partners by a linear supermultiplet.$^{7,8}$ The linear
formulation makes direct use of the Kalb-Ramond field, which
is very natural in view of the corresponding string vertex operator.

The linear multiplet $L$ corresponds to a real vector
multiplet of the form:
\begin{equation}
L = \{\,l,\chi,0,0,h^\mu,-\!\!\not\! \partial\chi,-\Box l\,\}\, ,
\label{l}\end{equation}
where $l$ is the dilaton, $\chi$ the dilatino, and $h^\mu$ is the dual
field strength of the antisymmetric tensor field $b_{\lambda\rho}$:
\vglue -5mm \begin{equation}\textstyle
h^\mu= \frac{1}{2}\epsilon^{\mu\nu\lambda\rho}\partial_{\nu}b_{\lambda\rho}.
\label{h}\end{equation}

The tree level Lagrangian defined by the gauge function
and the K\"ahler  potential of Eq.(\ref{tree}) is equivalent
to a Lagrangian constructed from the d-density\footnote{Here,
I ignore the subtleties related to the so-called chiral compensator
which I set $\Sigma_0=1$.} of $(L-2k\Omega )^{-1/2}
e^{-G^{(0)}/2}$, where $\Omega$ is the Chern-Simons vector superfield
(the vector component of $\Omega$ is the gauge topological current
$\omega^{\mu}$, $\partial_{\mu}\omega^{\mu}=F\widetilde{F}$).
This can be shown by performing a supersymmetric
generalization of the duality transformation between the Kalb-Ramond
field and the pseudoscalar axion. Note that although the Chern-Simons
field is not gauge invariant, the invariance of the Lagrangian is
ensured by the appropriate transformation property of $L$,
so that the combination $\widehat{L}=L-2k\Omega$ remains invariant.
A simple power counting argument$^{8,9}$ shows that
the string loop expansion must generate a d-density of the form:
\begin{equation}
{\rm d}= \widehat{L}^{-1/2}e^{-G^{(0)}/2} +\widehat{L}G^{(1)}
+{\cal O}(\widehat{L}^{5/2})\, .
\label{exp}\end{equation}
The one-loop term proportional to $\widehat{L}$ is usually called
the Green-Schwarz term$^{10}$ because it can be interpreted as the
compactification of the ten-dimensional term involved in the
Green-Schwarz anomaly cancellation mechanism.$^{11}$

In the linear formulation, the gauge function depends on the
chiral superfields $Z$ only, and by a similar argument
it consists entirely of the one loop contribution:\footnote{The tree level
gauge kinetic terms are contained in the d-density of Eq.(\ref{exp});
this will become clear in Eq.(\ref{delta}).} $f=f^{(1)}(Z)$.
On the other hand, the superpotential does not receive
any loop corrections: $W=W^{(0)}(Z)$, in agreement with the
well known supersymmetric non-renormalization theorems.
In the following, I will discuss the relation of
the one-loop functions $G^{(1)}$ and $f^{(1)}$ to the physical
parameters of superstring theory.

\vglue 0.4cm
{\elevenbf \noindent 3. Effective Action and
Physical Parameters in Superstring Theory \hfil}
\vglue 0.2cm
The tree-level effective action describes interactions of massless
string excitations at energies below the string scale.
These include contact interactions
due to the propagation of heavy particles in one-(massive)-particle
reducible diagrams. The masses of heavy particles depend on the
moduli, \eg\ the radii of compactified dimensions, therefore
the induced massless particle interactions are also moduli-dependent.
In order to compute the one-loop effective action, one should
integrate all diagrams involving heavy particles propagating
inside loops. In string theory, it is very difficult to separate
heavy from massless particles in higher genus diagrams, therefore
the computation of the effective action becomes a bit more
subtle than in field theory.
In order to explain the procedure employed in string theory, I
will discuss the computation of the one-loop terms that
determine the moduli-dependence of gauge couplings. These terms
are contained in the interaction of the form $-\frac{1}{4}\Delta
F_{\mu\nu}F^{\mu\nu}$, where $\Delta (l,z,\bar{z})$ is a real function
which plays the role of a field-dependent $1/g^2$,
$g$ being the gauge coupling constant.

In order to compute the function $\Delta$ one considers the two-point
correlation function of gauge bosons.$^{2,3,5}$
After insterting two gauge boson
vertices on a world-sheet torus one obtains the on-shell ($Q^2=0$)
correlation function as an integral over
the Teichm\"uller parameter $\tau$ of the torus. The integrand depends on the
moduli, reflecting moduli-dependent masses of heavy particles
propagating in the loop.
Since there are also massless particles
propagating in the loop, and the external particles are on-shell,
the integral over the Teichm\"uller parameter diverges in the infrared.
In the analogous background field-theoretical computations, such a
logarithmic divergence is usually regulated by going off-shell, to
momentum $Q^2\neq 0$. These logarithmic ($\ln Q^2$) terms play very important
role in the effective field theory -- they cause the ``running''
of gauge coupling constants.
It is very important to realize
that in string theory, as well as in quantum field theory,
the momentum-dependence of gauge coupling constants
is a purely infrared effect, therefore the corresponding
beta function coefficient of the $Q^2\rightarrow 0$ divergence
depends on the massless particle content only. The momentum-dependence
is governed by the usual renormalization group equations.
Hence, as far as running is
concerned, there is no difference between
string and standard unifications.$^{5,12}$

The basic difference  between string and standard unifications concerns
the boundary conditions at the unification scale.
Whereas in the standard case the boundary conditions are completely arbitrary,
in string theory they are determined by dynamical VEVs.
In the leading approximation, the tree-level part of the d-density
(\ref{exp}) gives $\Delta = k/l$ hence,
at the string unification scale $M$, $kg^2(M)=kg^2_{\rm tree}=l$
is equal for all
gauge couplings. At the one loop order, one expects however the so-called
threshold corrections which may significantly
affect the boundary conditions for the  next-to-leading
order renormalization group equations.$^{5}$

Let me go back now to the string computation of the one-loop
correlation function of two gauge bosons.
When the logarithmic divergence in the Teichm\"uller parameter
integration is regularized and compared to the field-theoretical
$\overline{D R}$ scheme, it is converted to $\ln(Q^2/M^2)$,
where $Q$ is the infrared cutoff.$^{5,12}$
The remaining finite part of the integral yields moduli-dependent
threshold corrections.

\underline{\it Example}. For the class of orbifolds mentioned in
Sec.{\it 2.1}.,$^{2,3}$\vglue -3mm
\begin{equation}
\Delta_{E_6}=\frac{1}{g^2_{E_6}(M)}=\frac{1}{g^2_{\rm tree}}-
\sum_{j=1}^3\hat{b}^j_{E_6}\ln [|\eta(iT_j)|^4(T_j+
\overline{ T}_{\bar{\jmath}})]+c_{E_6}\, ,
\label{exam}\end{equation}\vglue -2mm
\noindent where $\hat{b}^j_{E_6}$ and $c_{E_6}$ are
constants, and $\eta$ is the Dedekind eta function.

The moduli-dependence of threshold correction can be
determined also in a way that circumvents the problem
of infrared divergences. As a consequence of supersymmetry,
the derivatives of $\Delta$ with
respect to scalar fields are related to the couplings
of the corresponding pseudoscalars to gauge bosons.$^{2,3}$
The most useful relation is given by the so-called
integrability equation
$\partial_T\partial_{\overline{T}}\Delta=\frac{i}{2}
(\partial_{\overline{T}}\Theta_T
-\partial_T\Theta_{\overline{T}})$, where $\Theta_T$ is the $C\! P$-odd
part of the scattering amplitude involving  one modulus $T$ and two gauge
bosons. $\Theta_T$ is infrared finite.
$\partial_T\partial_{\overline{T}}\Delta$ have been computed
from this amplitude for the untwisted moduli in orbifold models.$^{2,3}$
{}From there, one can determine
$\Delta$, up to moduli-independent constants, by using
the invariance of superstring theory under the $SL(2,Z)$
duality transformations generated by $T\rightarrow 1/T$ and
$T\rightarrow T+i$.

The threshold corrections must be consistent with
the structure of one-loop supergravity action discussed in Sec.{\it 2.2.}
The gauge kinetic terms correspond to
\begin{equation}
\Delta = k\,[1/l+G^{(1)}]+{\rm Re}f^{(1)}.
\label{delta}\end{equation}
There is one subtlety involved in reading
the functions  $G^{(1)}$ and $f^{(1)}$ off $\Delta$ as determined from
scattering amplitudes. Eq.(\ref{delta})
was derived assuming locality of the effective action.
In fact, this assumption is violated due to one-loop anomalous
couplings of pseudoscalar axions to gauge bosons.
It is well known that such anomalous interactions are non-local;$^{13}$
they involve operators of the form $\Box^{-1}$.
The form of these additional terms is also known:
they can be determined on purely field-theoretical
grounds after one reads off the tree-level supergravity
Lagrangian the couplings of pseudoscalars to massless chiral
fermions.$^{10,14}$
Taking into account the presence of these extra terms
one can determine the moduli-dependence of the Green-Schwarz term
and from there, by using the $SL(2,Z)$ invariance,
the gauge function $f^{(1)}$.
The results$^{10,14}$ for the Green-Schwarz term agree with yet another
computation which will be discussed later on.

As far as higher order loops are concerned, the moduli-dependence
of radiative corrections to gauge couplings satisfies a
non-renormalization theorem.$^3$ In the case of orbifold models,
it is given entirely by the one-loop contributions; the
proof follows from a direct inspection of higher genus diagrams.
\vglue 0.4cm
{\elevenbf \noindent 4. Threshold Corrections to Yukawa Couplings
and K\"ahler metric\hfil}
\vglue 0.2cm

The one-loop supergravity action of Sec.{\it 2.2}.\ dictates the
following from of Yukawa interactions between chiral fermions $\psi$
and scalars $z$:
\begin{equation}\textstyle
{\cal L}_Y= -\frac{1}{2}\sqrt{l/2}\, e^{G^{(0)}/2}\, W^{(0)}_{ijk}\,
\psi^i\psi^j z^k + c.c.,
\label{ly}\end{equation}
where the subscripts of denote differentiation with respect to
the corresponding fields. Note that the above expressions depends
on the tree-level quantities only. The physical Yukawa couplings
defined by the fermion-scalar scattering amplitudes
may receive however loop corrections. They arise from the
wave the function renormalization factors, \ie\ from the corrections
to the K\"ahler metric.

In order to discuss loop corrections to the K\"ahler metric, it is
sufficient to consider the bosonic part of the kinetic energy terms:$^{7,8}$
\begin{equation}
{\cal L}_B =
- \frac{1}{4l^2}\partial_{\mu}l\partial^{\mu}l + \frac{1}{4l^2}
h_{\mu}h^{\mu} - G_{i\bar{\jmath}}\partial_{\mu}{z^i}\partial^{\mu}
{\bar{z}^{\bar{\jmath}}} -
\frac{i}{2}(G^{(1)}_{j}\partial_{\mu}{z^j} -
G^{(1)}_{\bar{\jmath}}\partial_{\mu}{\bar{z}^{\bar{\jmath}}})h^{\mu}.
\label{lb}\end{equation}
At the one-loop level, the K\"ahler metric is\vglue -2mm
\begin{equation}
G_{i\bar{\jmath}}=G^{(0)}_{i\bar{\jmath}}+lG^{(1)}_{i\bar{\jmath}}.
\label{kahler}\end{equation}

The main advantage of using linear formulation of
supergravity is that it provides a simple way of computing
loop corrections to the K\"ahler metric, by considering
a three-point amplitude involving two complex scalars and one Kalb-Ramond
field.$^{4}$
Inspection of the last term in Eq.(\ref{lb}) shows that
\begin{equation}
\epsilon^{\mu\nu\lambda\rho}p_{1\lambda}p_{2\rho}\,G^{(1)}_{i\bar{\jmath}}
=\langle z_i(p_1)\bar{z}_{\bar{\jmath}}(p_2)b_{\mu\nu}(p_3)\rangle .
\label{amp}\end{equation}
Although this amplitude vanishes for on-shell Minkowski momenta,
it can be computed for complex Euclidean momenta, as it was done in
similar computations for moduli and gauge bosons.$^2$ As a result,
one obtains:$^{4}$
\newcommand{\z}{\zeta}\newcommand{\zbar}{\bar{zeta}}
\begin{equation}\textstyle
G^{(1)}_{i\bar{\jmath}}=
\int\frac{d^2\tau}{{\tau_2}^2} \int
d^2{\z} \,\bar{\eta}(\bar{\tau})^{-2}\,
\langle\Psi_i(\z)\overline{\Psi}_{\bar{\jmath}}(0)\rangle ,
\label{gzz}\end{equation}
where $\Psi_i$ and $\overline{\Psi}_{\bar{\jmath}}$ are
the corresponding primary fields.

The integral over the Teichm\"uller parameter
in Eq.(\ref{gzz}) is infrared divergent. As in the case of
gauge couplings, these divergences are due to massless particles
propagating in the loop. The coefficients of divergent terms
correspond to the one-loop anomalous dimensions. They
can be extracted from Eq.(\ref{gzz}) which is valid for a general
compactification. The comparison with
field-theoretical anomalous dimensions shows that the string
computation implicitly uses a gauge in which the superpotential
remains unrenormalized. Again as in the case of gauge couplings,
the momentum-dependence of physical
Yukawa couplings in string theory
turns out as determined by the corresponding
field-theoretical beta functions.$^{15}$ The remaining finite
part of $G^{(1)}_{i\bar{\jmath}}$
gives the string threshold corrections to wave function factors.
These corrections determine the boundary conditions for the physical
Yukawa couplings $\lambda_{ijk}$ at the unification scale:
\begin{equation}
\lambda_{ijk}(M)=\lambda_{ijk}^{\rm tree}\, [1+l\, (Y_i+Y_j+Y_k)]^{-1/2},
\label{yukawa}\end{equation}
where $Y_i$ is defined as the finite part of $\displaystyle
G_{\mbox{}}^{(0)i\bar{\jmath}}G^{(1)}_{i\bar{\jmath}}$ (no summation
over $i$).

The one-loop K\"ahler metric and threshold corrections to Yukawa
couplings can be explicitly computed from Eq.(\ref{gzz})
in the case of orbifold compactifications. For the untwisted
moduli, $G^{(1)}_{T\overline{T}}$ obtained in this way
agrees with the corresponding contribution of the Green-Schwarz term to
the threshold corrections to gauge couplings
discussed in Sec.{\it 3}, \cf\ Eq.(\ref{delta}).

For the untwisted 27's and $\overline{27}$'s of $E_6$
one obtains non-vanishing wave function corrections only if the orbifold
group contains a subgroup that leaves invariant one of the three
complex orbifold planes and preserves $N=2$ supersymmetry.
The threshold correction $Y_z$ for an unwisted $z$-field
(27 or $\overline{27}$) associated
with such a plane depends on its moduli $T$, and does {\em not\/}
depend on the moduli of other planes:
\begin{equation}
Y_z=\frac{2\hat{\gamma}_z}{ind}\,\ln [|\eta (iT)|^4(T+\overline{T})]+y_z\, ,
\label{y}\end{equation}
Here, the coefficients $\hat{\gamma}_z$ are anomalous dimensions of
the $z$-fields in the corresponding $N=2$ supersymmetric theory
with the orbifold defined by the little group of the unrotated
plane associated with the modulus $T$, and $ind$ is the index
of this little group in the full orbifold group;
$y_z$ are moduli-independent constants. One can show that
$\hat{\gamma}_z=-\hat{\beta}_z/2$, where $\hat{\beta}_z/2$ is
the corresponding beta function coefficient of any gauge subgroup
that transforms $z$ non-trivially in the embedding $N=2$ theory.
The threshold corrections to Yukawa couplings can be computed
by substituting Eq.(\ref{y}) into Eq.(\ref{yukawa}).

\underline{\em Example}. As an example, consider the Yukawa
coupling between three untwisted 27's of $E_6$. At the tree-level,
this coupling is:
\begin{equation}
\lambda_{ijk}^{\rm tree}=\frac{g_{\rm tree}}{\sqrt{2}}W_{ijk}\, ,
\end{equation}
where $W_{ijk}$ are constants which are non-zero only if
the three 27's are associated with three different planes.
In this case, Eq.(\ref{y}) combined with Eq.(\ref{yukawa})
give the boundary condition:$^4$
\begin{equation}
\lambda_{ijk}(M)=\frac{g_{E6}(M)}{\sqrt{2}}W_{ijk}\,
[1+g^2_{E6}(M)y_{ijk}]^{-1/2},
\end{equation}
where $y_{ijk}$ are moduli-independent constants, and
$g_{E6}(M)$ is the one-loop $E_6$ gauge coupling at the
unification scale, \cf\ Eq.(\ref{exam}).
As a result, the boundary relation between the untwisted
Yukawa couplings and the $E_6$ gauge coupling at the
unification scale does not receive any moduli-dependent
corrections at the one-loop level.
\vglue 0.4cm
{\elevenbf\noindent 5. Conclusions \hfil}
\vglue 0.2cm

It is a common opinion among high energy physicists that superstring
theory is still far from becoming the theory of everything.
There remain many fundamental problems to be solved before
a reliable low-energy phenomenology can be developed.
What remains however unquestionable is that superstring
theory provides a unique example of a perfectly consistent
unification of gravitational and gauge interactions, within
the framework of local supersymmetry. The efforts described
here have been motivated by the desire to understand the
low-energy limit of such a consistent theory. As a result,
we have now a very good understanding of the low-energy physics
of supergravity theory that describes not
only the classical limit of string theory, but also some interesting
string loop effects, in particular the threshold corrections.
This ``bottom--up'' approach to superstring theory may well provide
some new insights into the Planck-scale physics.
\vglue 0.4cm
{\elevenbf\noindent Acknowledgements \hfil}
\vglue 0.2cm

It is my great pleasure to thank I. Antoniadis,
E. Gava and K.S. Narain for enjoyable collaborations. This work
was supported in part by the Northeastern University Research and
Scholarship Fund and in part by the National Science Foundation
under grant PHY-91-07809.
\vglue 0.4cm
{\elevenbf\noindent References \hfil}
\vglue 0.2cm

\end{document}